\documentclass[aps,prl,reprint,twocolumn,groupedaddress]{revtex4}
\usepackage{graphicx}
\usepackage{dcolumn}
\usepackage{bm}
\usepackage{braket}
\begin{document}
\title{Non-Abelian $\nu=1/2$ quantum Hall state in $\Gamma_8$ Valence Band Hole Liquid }
\author{George Simion}
\email{simion@purdue.edu} \affiliation{Department of Physics and Astronomy and  Purdue Quantum Center, Purdue University, West Lafayette IN, 47907 USA}
\author{Yuli Lyanda-Geller}
\email{yuli@purdue.edu} \affiliation{Department of Physics and Astronomy and Purdue Quantum Center, Purdue University, West Lafayette IN, 47907 USA}

\date{March 10, 2016}

\begin{abstract}
In search of states with non-Abelian statistics, we explore the fractional quantum Hall effect in a system of two-dimensional charge carrier holes. We propose a new method of mapping states of holes confined to a finite width quantum well in a perpendicular magnetic field to states in a spherical shell geometry.
This method provides single-particle hole states used in exact diagonalization of systems
with a small number of holes in the presence of Coulomb interactions. An incompressible fractional quantum 
Hall state emerges in a hole liquid at
the half-filling of the ground state in a magnetic field in the range of fields where single-hole states cross. This state has a negligible overlap with the Halperin 331 state, but a significant overlap with the Moore-Read Pfaffian state. Excited fractional quantum Hall states for small systems have sizable overlap with non-Abelian excitations of the Moore-Read Pfaffian state.

\end{abstract}

\maketitle

Quasiparticles obeying non-Abelian statistics lead to fault tolerant quantum computing
\cite{NayakDasSarmaRPMQC, BonesteelSimonPRL, DasSarmaNAPRL}. Exotic states resulting in non-Abelian excitations
can arise in low dimensional quantum liquids in the
presence of magnetic fields.  The fractional quantum
Hall (FQH) state in a two-dimensional (2D) electron liquid at filling factor $\nu=5/2$  is the state most studied theoretically and
possibly observed experimentally \cite{MooreRead, ReadGreen, Review, Willett52}. There are
 other FQH states at $\nu=12/5$
\cite{ReadRezayiPRB99} and $\nu=8/3$ \cite{ReadRezayiPRB99,
SpinSingletPRB, BarkeshliWenNATPRL}, bilayer $\nu=1/2$ 2D electron phase\cite{ShayeganPRL92,
ShayeganWinklerelPRB, PetersonDasSarmaPRB, PapicGoerbig12PRB}, and
$\nu=1/4$ state \cite{PanPRL14, PapicPRB14}, for which
non-Abelian origin of excitations has been discussed. Other candidates for non-Abelian systems are vortices in $p$-wave
superconductors \cite{Volovik}, and hybrid systems with proximity-induced $s$-wave superconductivity that
 mimic a $p$-wave pairing in semiconductors and topological insulators, due to spin-orbit coupling
\cite{DasSarmaMajoranaPRL10,Oreg10}, Dirac spectra
\cite{FuKanePRLMajorana}, or Laughlin anyon quasiparticles
\cite{Shtengel}.

Here we show that FQHE in 2D hole systems is a new
promising non-Abelian setting. Luttinger valence band holes are fundamentally
different from electrons. They exhibit non-Abelian phases in
transport even for single-hole states
\cite{ArovasYLGPRB}. In a magnetic field, the single-hole states
are four-component spinors. Each spinor component is described by a
distinct Landau level (LL) wavefunction $u_n$, $n\ge 0$. The relative
weights of these functions in spinors vary with magnetic
field \cite{SLGHoles}. Functions $u_1$ generating FQH
non-Laughlin electron correlations \cite{SimionQuinnPE} have
sizable weight in several hole states. Furthermore, the hole ground state
 in certain ranges of magnetic field is not defined by
$u_0$, like for electrons, but by $u_{n\ne0}$, including
$u_1$.  Thus,
 the non-Abelian FQH hole states can arise when the ground level in a single quantum well is filled.

Compared to electrons, holes have smaller cyclotron energy and stronger LL mixing by Coulomb interactions. Single-hole magnetic spectra exhibit multiple level crossings,  particularly in the ground state. Near crossings, ratio of interaction and cyclotron energies changes significantly for relatively small changes in magnetic field, and interaction pseudopotentials can be easily controlled.
A strong overlap, like for electrons at $\nu=5/2$ \cite{WojsJain}, is then possible with
Moore-Read \cite{MooreRead} or anti-Pfaffian states
\cite{HalperinPRL07}. Control of LL mixing was discussed
for crossing of electron levels dominated by $u_0$ and $u_1$, when
$\nu=2/5$ electron liquid is tuned by a small
change in magnetic field from a Laughlin state to a state with
non-Laughlin correlations and non-Abelian excitations \cite{BonesteelSimonPRL}. However, such electron cases are rare. Spectral crossings for holes are numerous, which makes the phase diagram for hole liquid much richer than that for electrons.

\begin{figure}
\centering
\includegraphics[width=0.45\columnwidth]{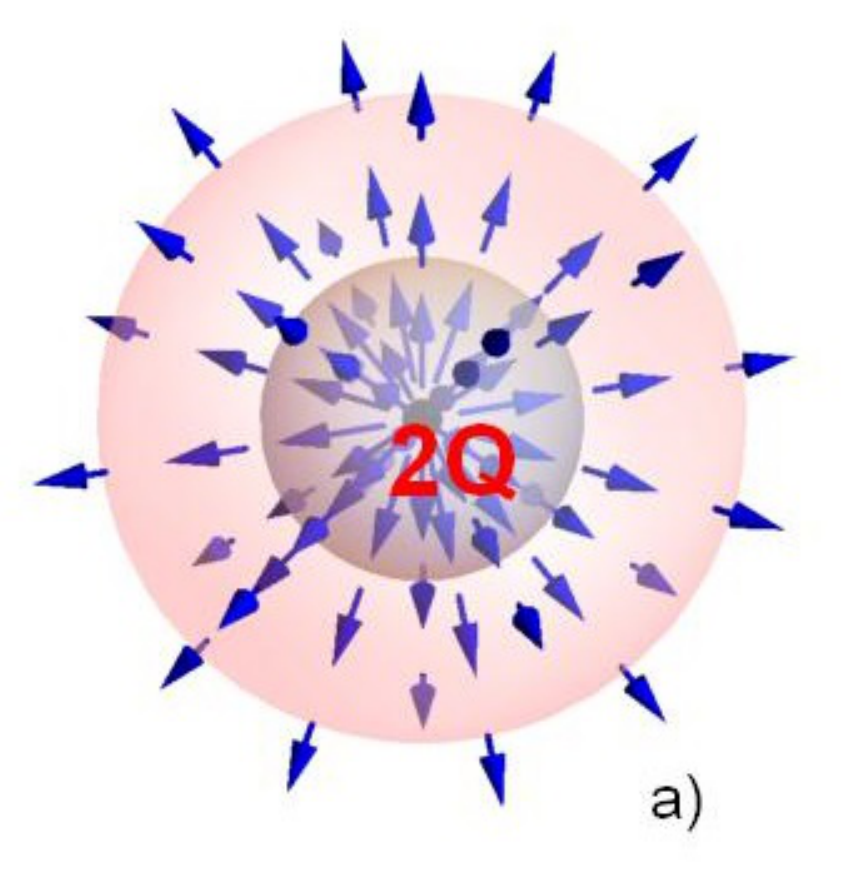}
\includegraphics[width=0.45\columnwidth]{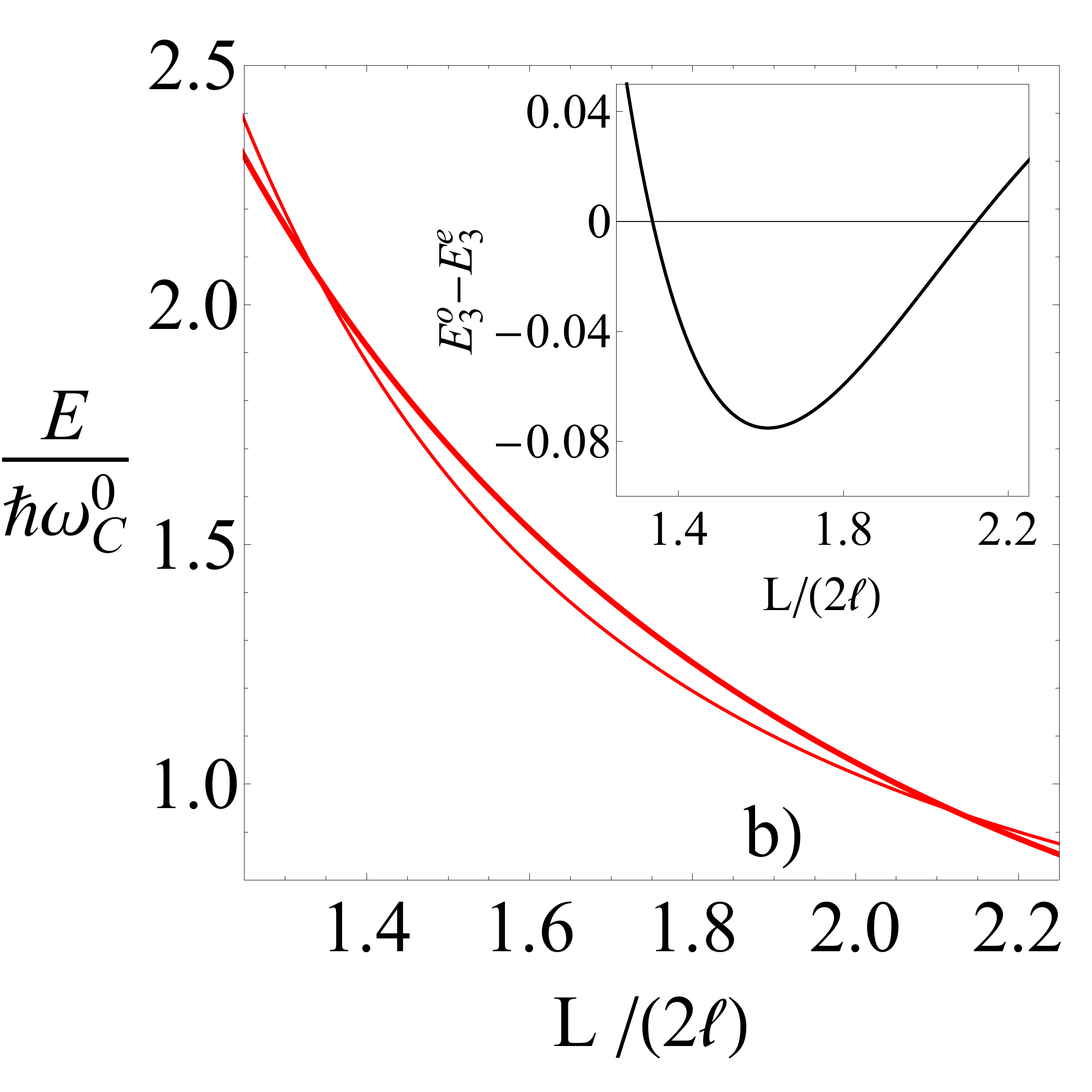}\\
\includegraphics[width=0.45\columnwidth]{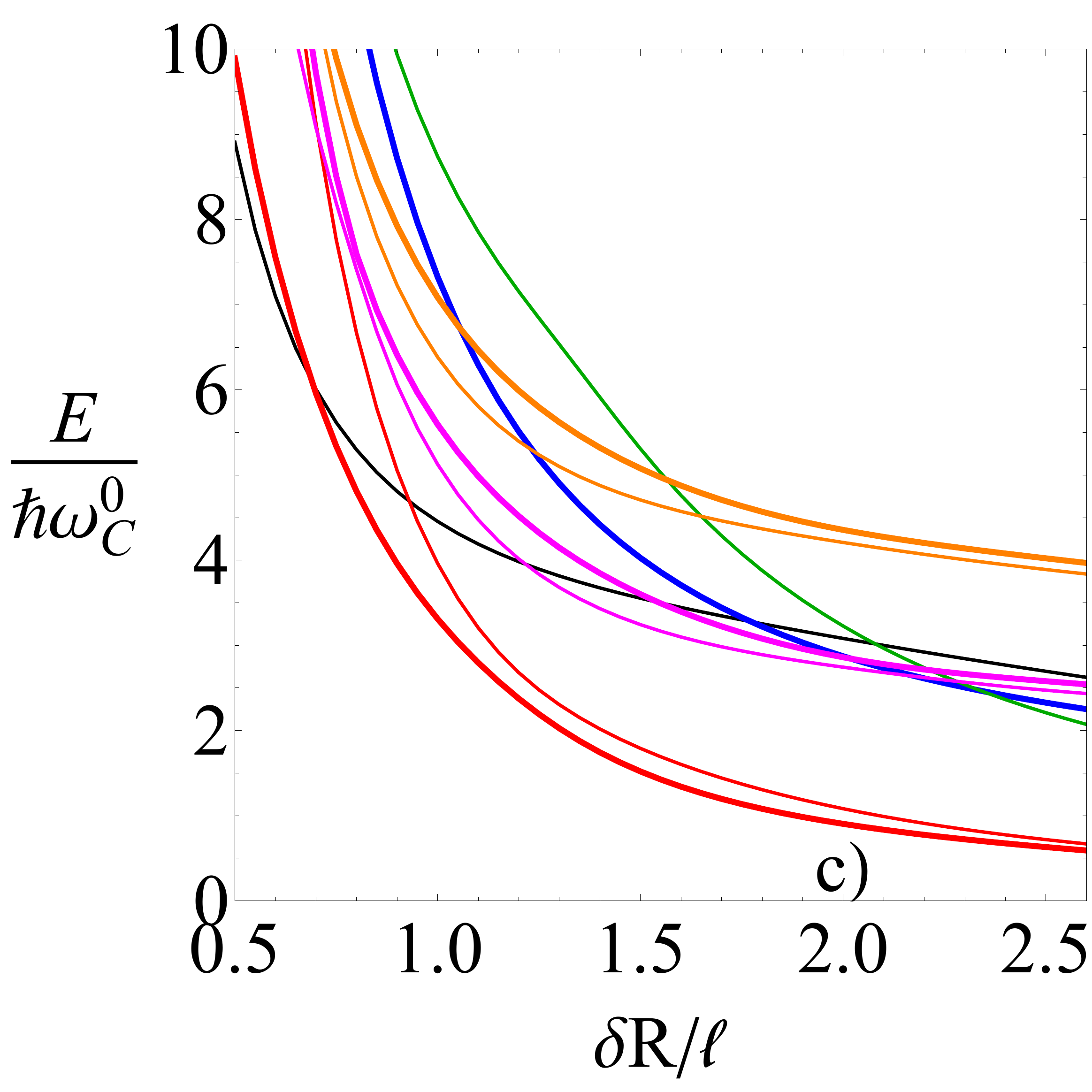}
\includegraphics[width=0.45\columnwidth]{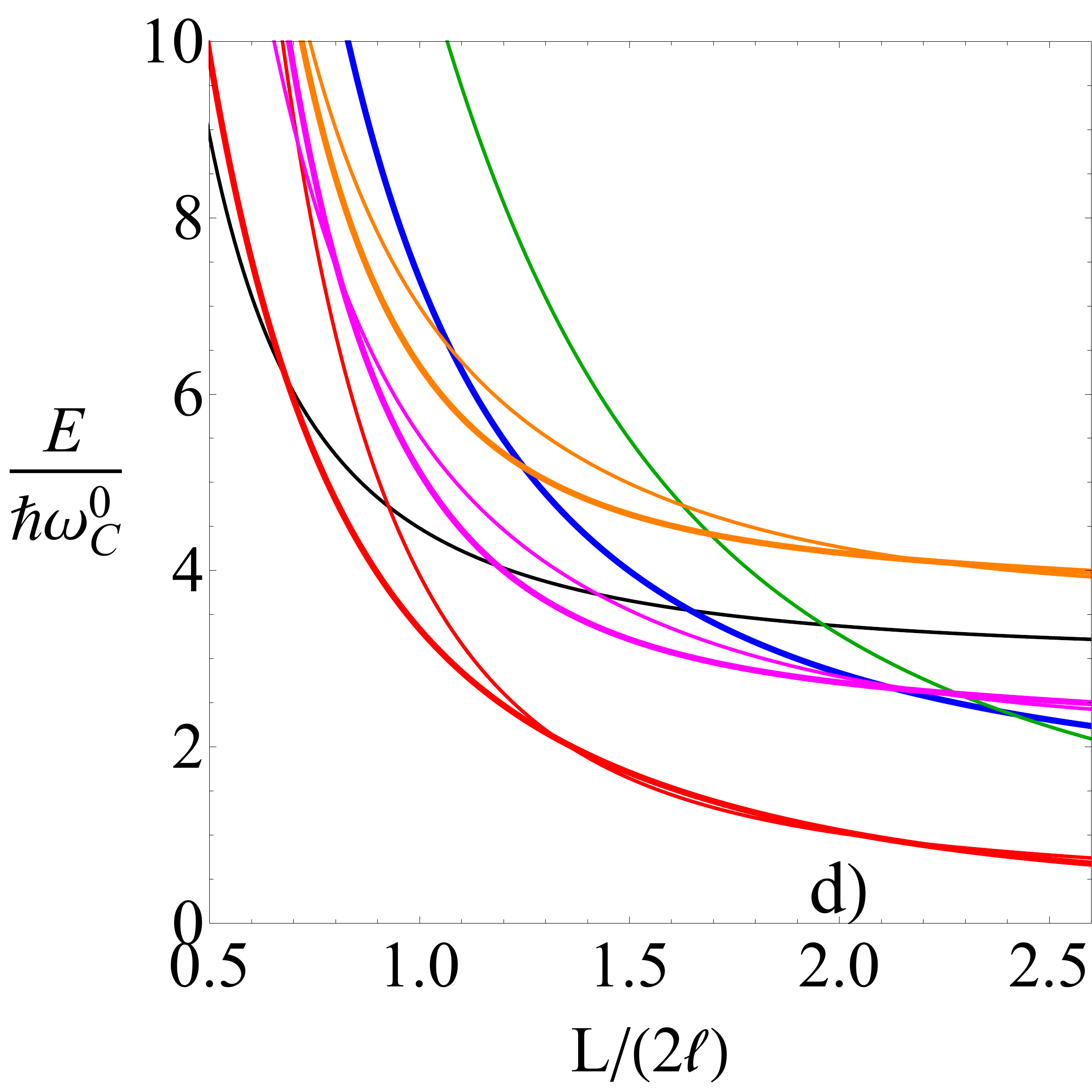}
\caption{Color online: a - Spherical shell geometry;  b - Ground state level crossings in a spherical 
shell (red solid lines) and planar geometry (black
dotted lines); c,d -
lowest nine states spectra ($n \leq 5$) in a spherical
shell geometry with $Q=100$ (c) and planar geometry (d).
The highest index Landau wavefunction in four-spinors of the shown hole states:
 Black lines - $u_0$; blue - $u_1$; green - $u_2$; red -
$u_3$; magenta- $u_4$; orange- $u_5$. Thick lines- even states, thin
lines-odd states. The thin red line state spinor has significant $u_1$-component.}\label{fig:gen_en_spectrum}
\vspace{-0.5cm}
\end{figure}

In search of non-Abelian hole states we propose a theoretical framework for treatment of FQHE  in hole systems.
Unusual hole spectra in magnetic field arise from strong coupling between the
 in-plane and spatial quantization $z-$direction motion in a quantum well, caused by strong spin-orbit interactions.
Hole four-spinors and the inseparability of the
in-plane and $z-$direction degrees of freedom make the treatment of
Coulomb interactions challenging. For electrons, the in-plane
and z-direction motion are independent, so it is possible to use the
Haldane technique\cite{HaldanePRL83} of homogeneous states with
translationally invariant wavefunctions for a finite number of
electrons on a sphere in a monopole magnetic
field. This method cannot be applied to holes. We propose a new
method of mapping quantum well confined hole states in a spherical
shell geometry, Fig. \ref{fig:gen_en_spectrum}. We then use our
method for consideration of the $\nu=1/2$ hole state at quantum well
widths corresponding to the range of magnetic fields with ground hole state crossings. We demonstrate that the FQH
state at $\nu=1/2$ is not the Halperin 331 state \cite{HalperinHPA} but rather a
Moore-Read (MR) state.

{\it {Holes in a planar and spherical shell geometry.}} The Luttinger Hamiltonian \cite{Luttinger56} in magnetic field $\mathbf{B}$ is
\begin{eqnarray}
\label{eq:Luttinger_simp} \hat
H_0&=&\left(\gamma_1+\frac{5}{2}\gamma\right)
\frac{\hat{\bf{k}}^2}{2}I-\gamma\left(\hat {\bf{k}} \cdot
{\bf{s}}\right)^2-\left(\frac{\gamma}{2}+\kappa\right) s_z,
\end{eqnarray}
where energies are in units of a free electron
cyclotron energy $\hbar\omega_c^0=\hbar e B/m_0 c$, dimensionless coordinates $\mathbf{r}$ are in units of magnetic length $(\ell=\sqrt{\hbar c/e B})$,
wavevectors  ${\bf{k}}=-i{\bf{\nabla}}_{\mathbf{r}} +e\ell {\bf{A}}/ (\hbar c)$, ${\bf{A}}$ is the vector potential, ${\bf{s}}$ is
spin $3/2$ operator, and $\gamma_1$, $\gamma$ and
$\kappa$ are Luttinger parameters in a spherical
approximation.
This Hamiltonian commutes with
the z-projection of total angular momentum $j_z=l_z+s_z$, $l$ is
the angular momentum.  In a symmetric gauge, the hole wavefunctions
 in a quantum well of width $L$ are:
\begin{equation}
\label{eq:basis_wf} \Psi_{n,m}^{\{\alpha\}}=\left(
\begin{array}{l}
\zeta_{0}^{\{\alpha\}}(z)u_{n,m}\\
\zeta_{1}^{\{\alpha\}}(z)u_{n-1,m+1}\\
\zeta_{2}^{\{\alpha\}}(z)u_{n-2,m+2}\\
\zeta_{3}^{\{\alpha\}}(z)u_{n-3,m+3}
\end{array}
\right)~,
\end{equation}
where $u_{n,m}$ are symmetric gauge eigenfunctions\cite{LLQM}, and $\zeta(z)$ are envelope functions
satisfying the boundary conditions $\Psi(\pm L/2)=0$. These wavefunctions reflect the correlation of the in-plane and $z-$direction motion, leading to a mutual transformation of heavy and light holes at the heterointerfaces due to giant spin-orbit coupling. Energies and wavefunctions are characterized by a single length
scale $w=L/(2\lambda)$\cite{SLGHoles}. For $n<3$, the components of wavefunctions
with $n-l<0$, $l=1,2,3$ vanish, and $n+1$ components are nonzero. The wavefunctions are even or odd in respect to
reflection about a plane $z=0$.

In order to construct  homogeneous states with translationally
invariant wavefunctions, we confine holes to a spherical shell with
radius $R_0-\delta_R\leq r \leq R_0+\delta_R$ as shown schematically
in Fig. \ref{fig:gen_en_spectrum} a. A magnetic field $B= 2Qhc/(4
\pi e r^2) $, is related to an integer monopole of strength $2Q$, so
that magnetic flux through spherical surfaces around it $\phi=
2Qh c/e$. Because ${\bf j}= {\bf{ l}}+{\bf s} $ is a good quantum
number for single-hole states, the eigenfunctions of
(\ref{eq:Luttinger_simp}) for a spherical shell are
\begin{eqnarray}
\label{eq:spherewf} &\psi_{\alpha jm}(r,\theta,
\phi)=\sum_{l=j-\frac{3}{2}}^{l=j+\frac{3}{2}}R_{\alpha j}^l(r)\times\nonumber\\
&\left(
\begin{array}{c}
\left<j,m\right|\left.l,m-\frac{3}{2};\frac{3}{2},+\frac{3}{2}\right> Y_{Q,l,m-\frac{3}{2}}(\theta, \phi)\\
\left<j,m\right|\left.l,m-\frac{1}{2};\frac{3}{2},+\frac{1}{2}\right> Y_{Q,l,m-\frac{1}{2}}(\theta, \phi)\\
\left<j,m\right|\left.l,m+\frac{1}{2};\frac{3}{2},-\frac{1}{2}\right> Y_{Q,l,m+\frac{1}{2}}(\theta, \phi)\\
\left<j,m\right|\left.l,m+\frac{3}{2};\frac{3}{2},-\frac{3}{2}\right>
Y_{Q,l,m+\frac{3}{2}}(\theta, \phi)
\end{array}
\right)~,
\end{eqnarray}
where $\left<j,m_j\right| \left.l,m-{l};\frac{3}{2},m_s\right>$ are
the Clebsch-Gordan coefficients of $\mathcal
{\bf{j}}={\bf{l}}+{\bf{s}}$, $Y_{Q,l,m}$ are the monopole harmonics
\cite{MonopoleHarmonics1}, and $\alpha$ labels subbands. Radial
functions $R_{\alpha j}^l(r)$ are defined by the boundary conditions $\psi_{\alpha
jm}(R_0\pm\delta_R)=0$.
Each wavefunction (\ref{eq:spherewf}) contains up to four spinors, each spinor having four
components. The monopole harmonics are defined if $l\geq Q$
\cite{MonopoleHarmonics1}, so that $2j\geq 2Q-3$. For
the states with $2j<2Q+3$, $j-Q+5/2$  spinor components are nonzero, while for
$2j\geq 2Q+3$ all components of spinors are non-zero. In Figs.
\ref{fig:gen_en_spectrum} c and d we present hole spectra
 in spherical and planar geometries. Each band of states
includes two states for every $2j\geq (2Q+3)$.

\begin{figure}
\centering
\includegraphics[width=0.45\columnwidth]{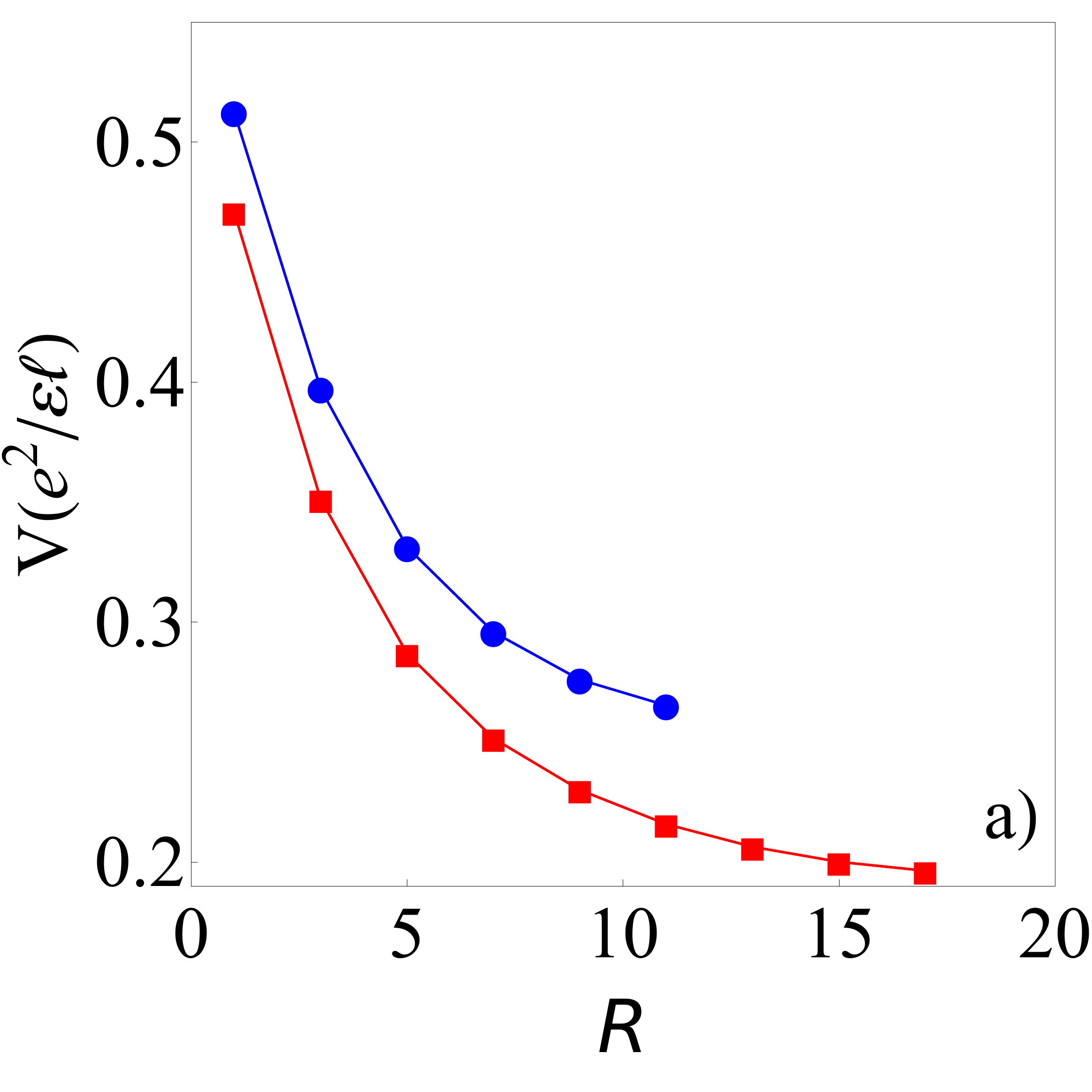},
\includegraphics[width=0.45\columnwidth]{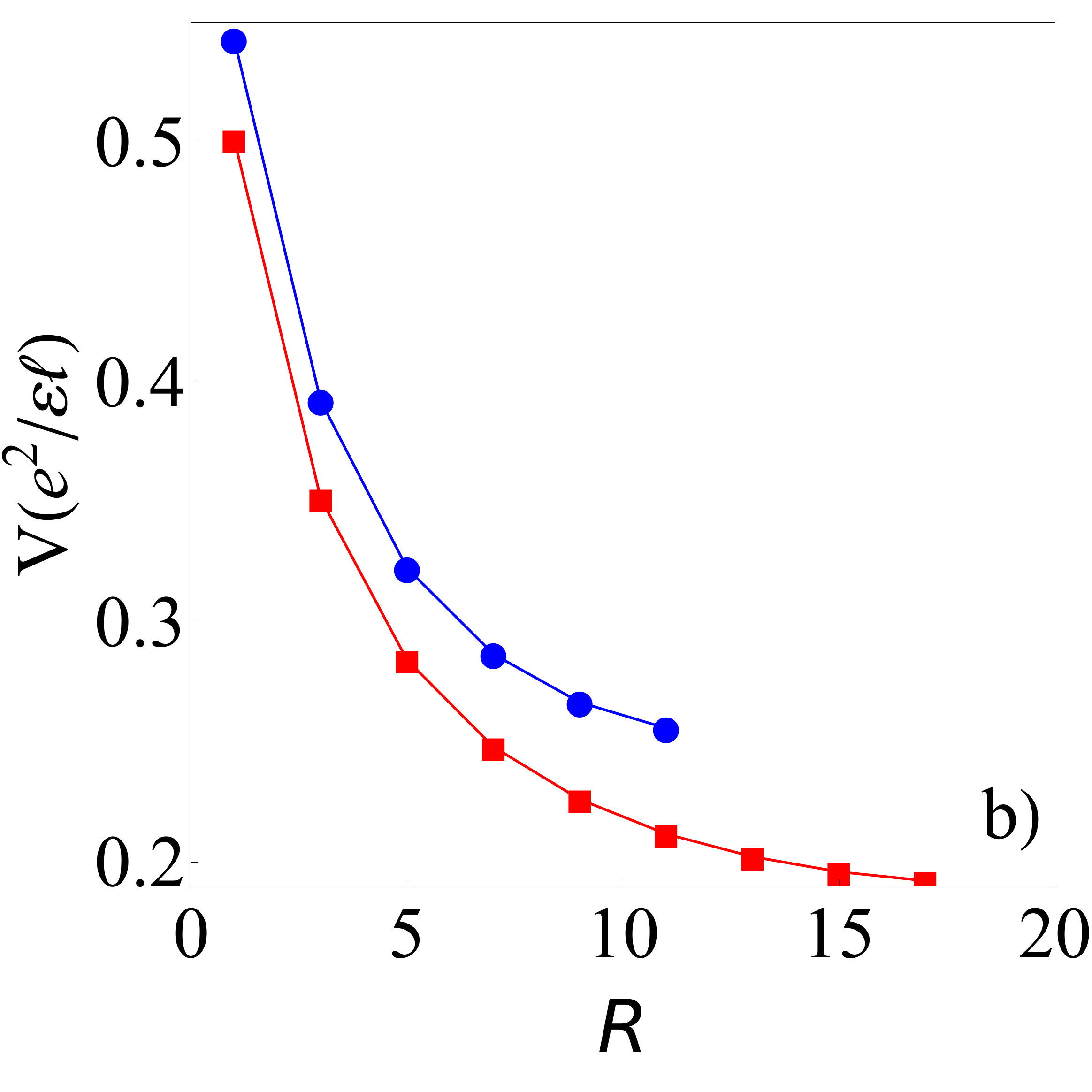}\\
\includegraphics[width=0.45\columnwidth]{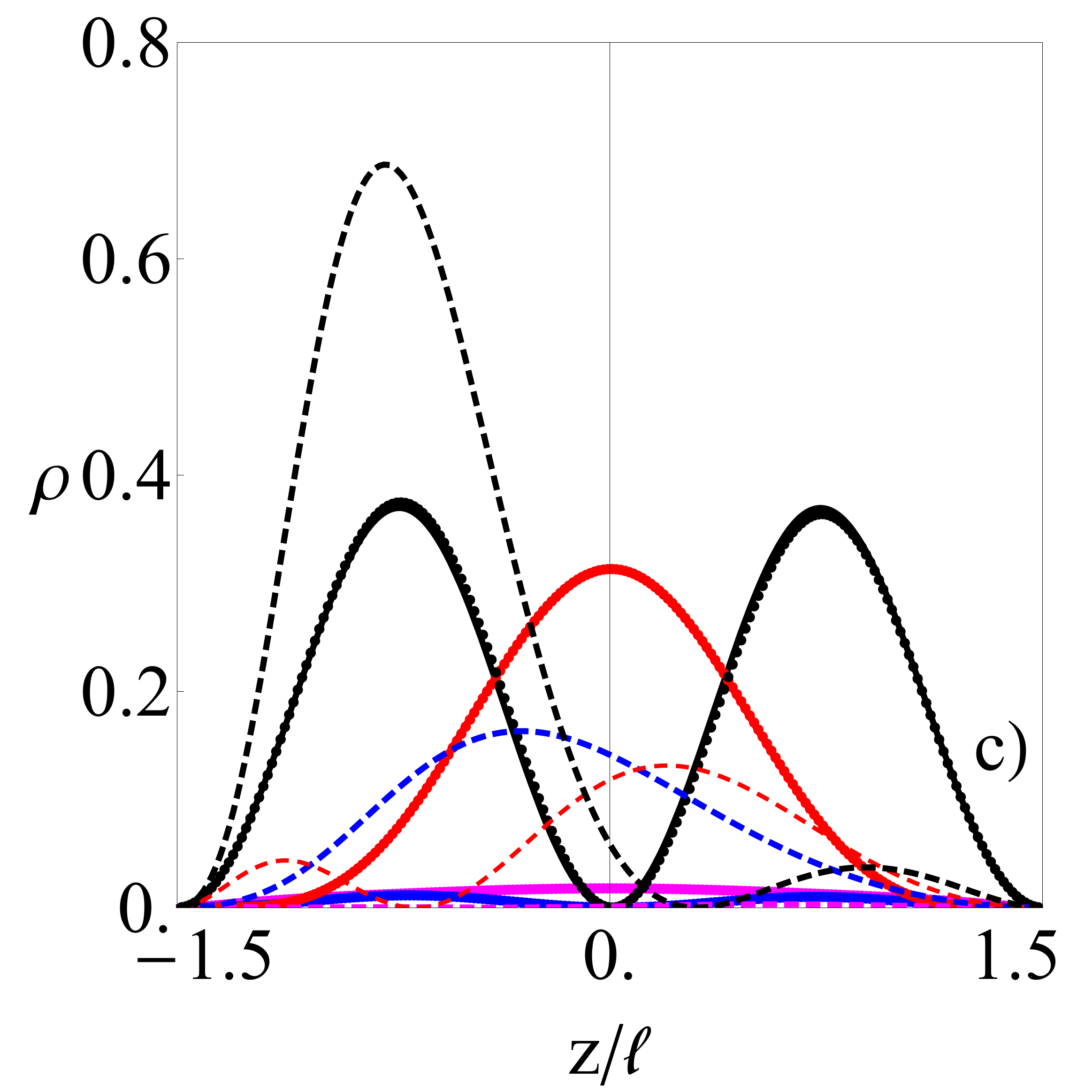},
\includegraphics[width=0.45\columnwidth]{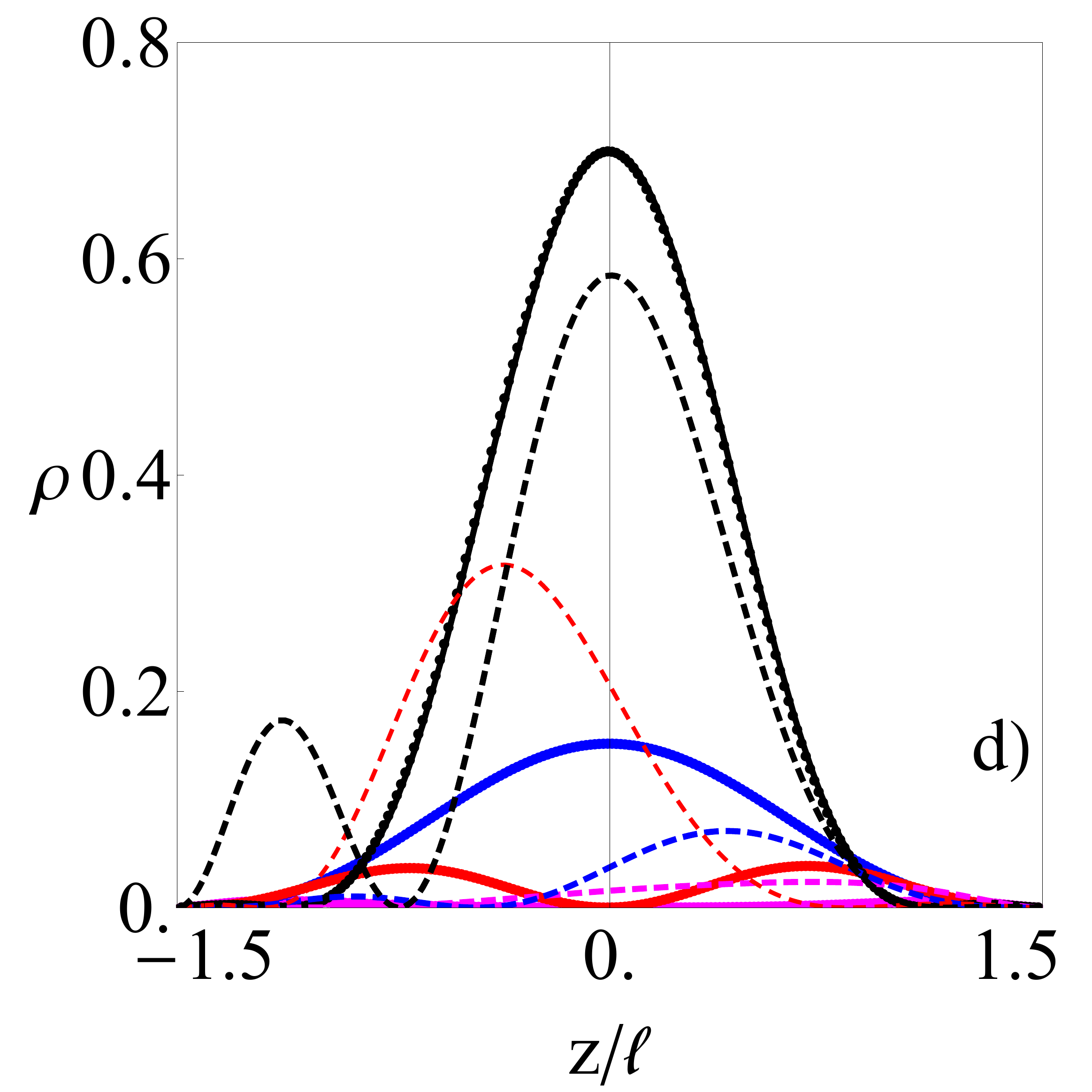}
\caption{Color online. a,b: pseudopotentials for $w=1.6$ for $2Q=10$ (blue dots) and
$2Q=15$ (red squares) for odd $n=3$ state (a), and even $n=3$ state (b).
c,d:  The charge density $\rho$. Vertical axis is for odd
$n=3$ state (c) and for even $n=3$ state (d). Black line: $-3/2$ spin
component (containing $u_0(r)$), red line: spin $-1/2$ spin component
(containing $u_1(r)$), magenta: 1/2 spin  component (containing $u_2
(r)$) blue: spin $3/2$ (containing $u_3 (r)$). The odd state has a
bigger $u_1$ admixture and its pseudopotential resembles that of LL1 
electrons, while the even state pseudopotential resembles that of LL0 electrons. In c and d, dashed lines correspond to $Q=15$, dotted lines
represent $Q=10^8$ and solid lines are for the planar
case.}
\label{fig:psedopot}
\vspace{-0.2cm}
\end{figure}

The energy spectra in the planar and spherical shell geometry are
almost identical, and crossings of the corresponding states in both
geometries occur almost at the same ratio $w$. Energies  in a
spherical shell converge to the planar limit for very large $Q$ in
much the same way as the Haldane electron wavefunctions on the
sphere converge to their planar limit.  We note that for finite $Q$,
there is no even-odd reflection parity, but it is restored in the
large $Q$ limit. Fig. \ref{fig:psedopot} c-d, shows radial
distributions of charge for the lowest states of the quantum well. The radial distribution of charge
density converges to the planar limit at large $Q$.
Thus, mapping of quantum well holes over a spherical shell provides
one to one correspondence between states. Each spherical state with
total angular momentum $j$ corresponds to a planar state
characterized by index $n=j-Q+3/2$. Each spinor of spherical
wavefunctions with angular momentum $l$ corresponds to a spin
component in the planar geometry with $s_z=j-l$, and the radial
wavefunctions are spherical equivalents of the $z$-envelope
functions of the planar geometry.

Crucially, hole states mix different functions $u_{n}(r)$. Fig.
\ref{fig:psedopot}d shows that $u_0$ 
 favoring Laughlin electron correlations dominates $n=3$ even hole planar state 
and its spherical counterparts. However, $u_1$, favoring non-Laughlin correlations and non-Abelian excitations, is prominent in other hole states.
The weights of these functions for hole spinors depend on $w$, and can be
changed significantly
by a slight change of a magnetic field.

{\it{Coulomb Interactions}.} A system of identical charged particles
in a magnetic field is highly degenerate. The Coulomb interaction 
cannot be treated perturbatively.
Such systems are modeled using a small number of particles. We perform
exact diagonalization of Coulomb interactions
$H_i=\sum_{ij}\frac{e^2}{\epsilon r_{ij}}$ for holes in a spherical
shell geometry and discuss extrapolation to the thermodynamic
limit.
The single-particle Hilbert space is
defined by states (\ref{eq:spherewf}).
The many-body basis set is given by all wavefunctions obtained when
 $N$ holes are placed in single-particle states. We calculate the
Coulomb interactions matrix elements  using addition of angular
momenta. Their explicit expressions and a system of differential equations for radial components
of wavefunctions are presented in the Supplementary Material.

The integral of motion in our many-body system is the total angular
momentum $\bf{J}=\sum_{i}\bf{j}_i$ and its $z$-projection. We apply the
 Wigner-Eckart theorem \cite{edmonds1996angular}
\begin{equation}
<J', M', \beta'|H_i|J, M, \beta
>=\delta_{JJ'}\delta_{MM'}V_{\beta\beta'}(\mathcal J)~,
\end{equation}
and reduce the Hilbert space, by using
independence of interaction matrix elements on the z-projection of the total angular momentum of all holes, $J_z$.
Here index $\beta$ labels the multiplets of many-body states with the same total $J$ and the same total $M$, and
$V_{\beta\beta'}(J)=< J',\beta '|H_i| J,\beta>$ are the
pseudopotentials \cite{HaldanePRL83}. We first compute the principal contribution to the two-body pseudopotentials of two holes, each with
an angular momentum $j$, without including any virtual transitions to other states,
$V^{0}_{00}({\boldsymbol{\mathcal
J}}={\bf{j}}+{\bf{j}})\equiv V_0(\mathcal{R})$, where $\mathcal{R}=j_1+j_2- J$ is the 
relative angular momentum. For the two-body interactions, there is one
multiplet for each allowed value of $\mathcal J$. The two-hole pseudopotentials $V_0 (\mathcal {R})$
are shown in  Fig. \ref{fig:psedopot}a-b for holes whose wavefunctions are the spherical counterparts of the odd parity  $n=3$
planar state, and the even parity $n=3$
planar state, correspondingly. 

\begin{figure}
\centering
\includegraphics[width=0.45\columnwidth]{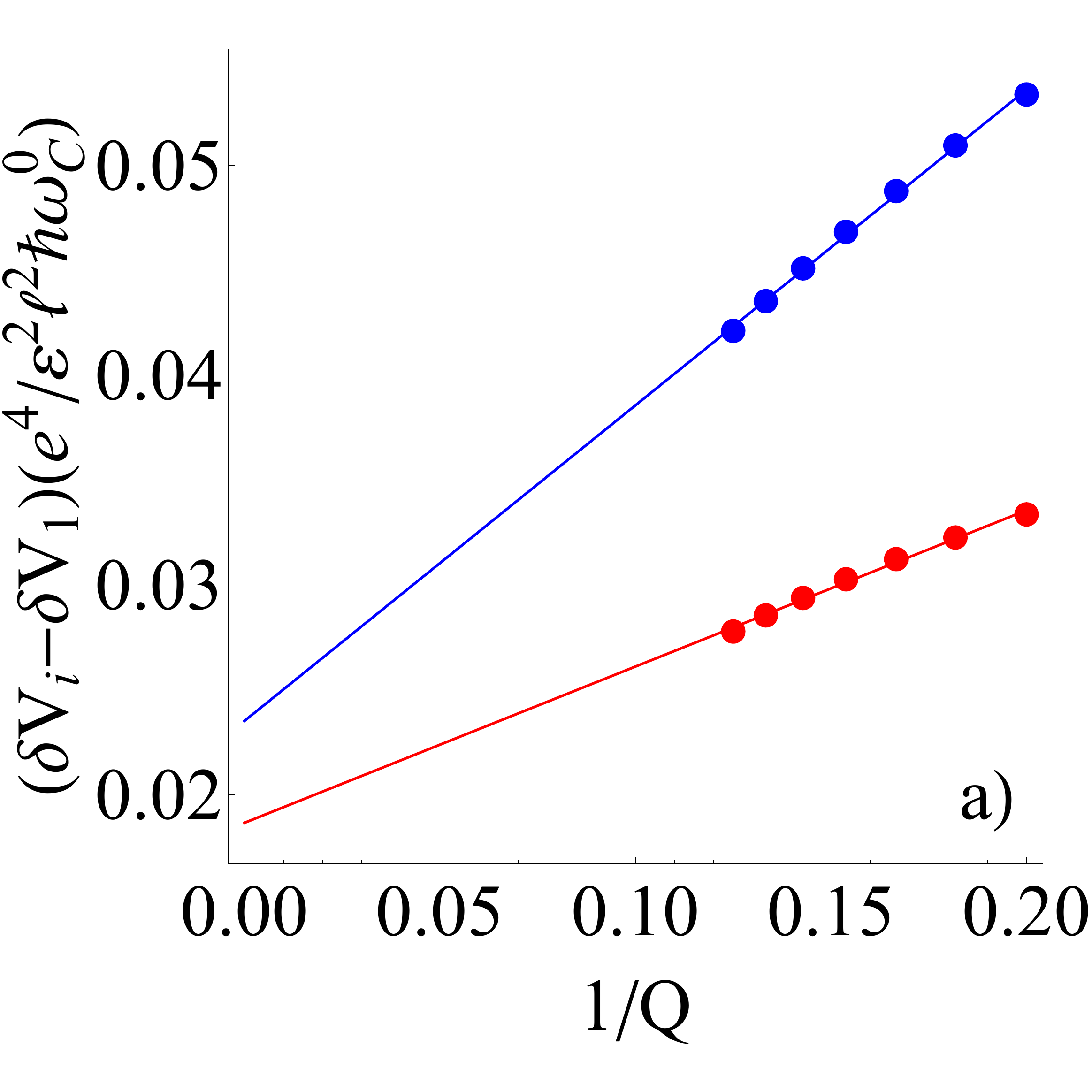},
\includegraphics[width=0.45\columnwidth]{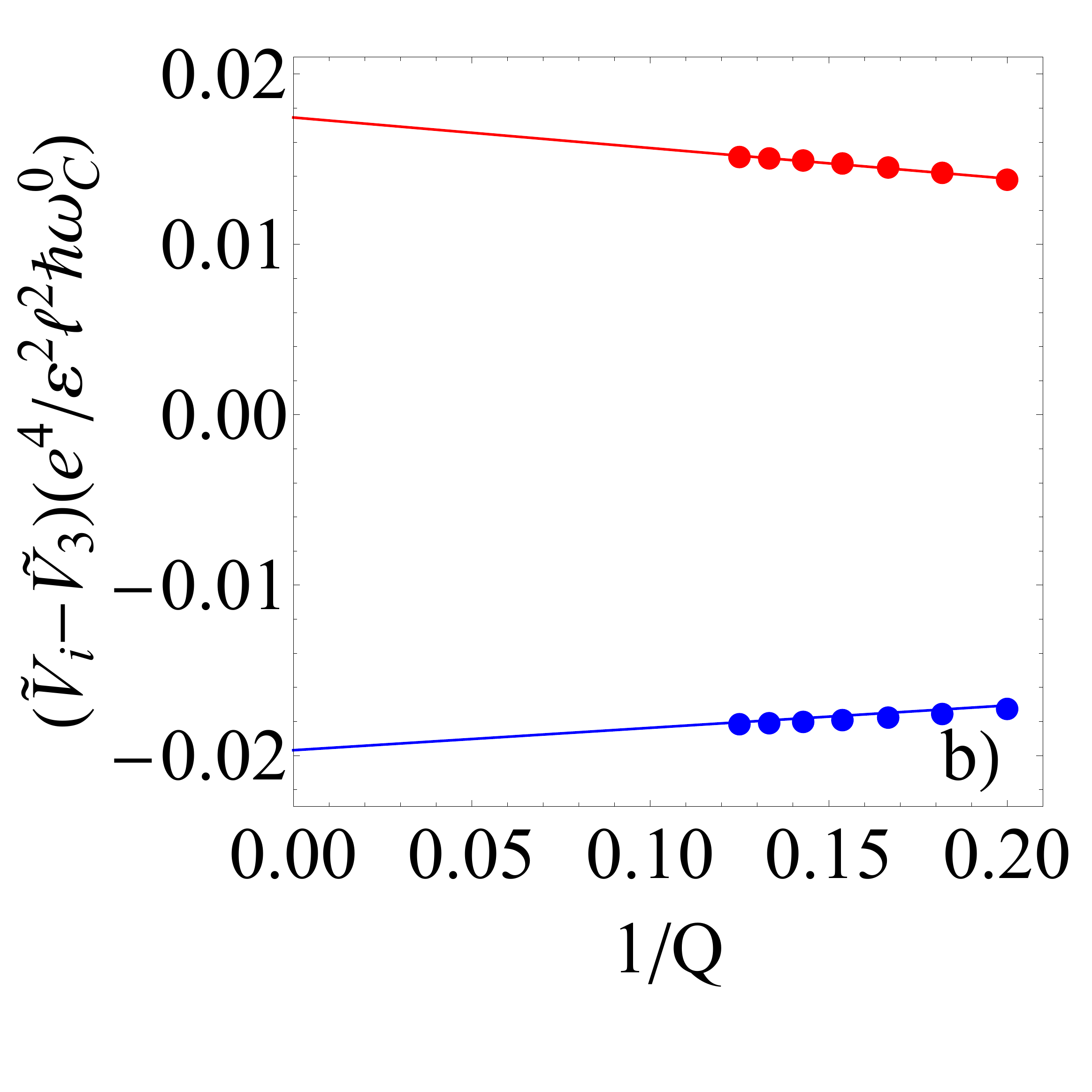}\\
\includegraphics[width=0.45\columnwidth]{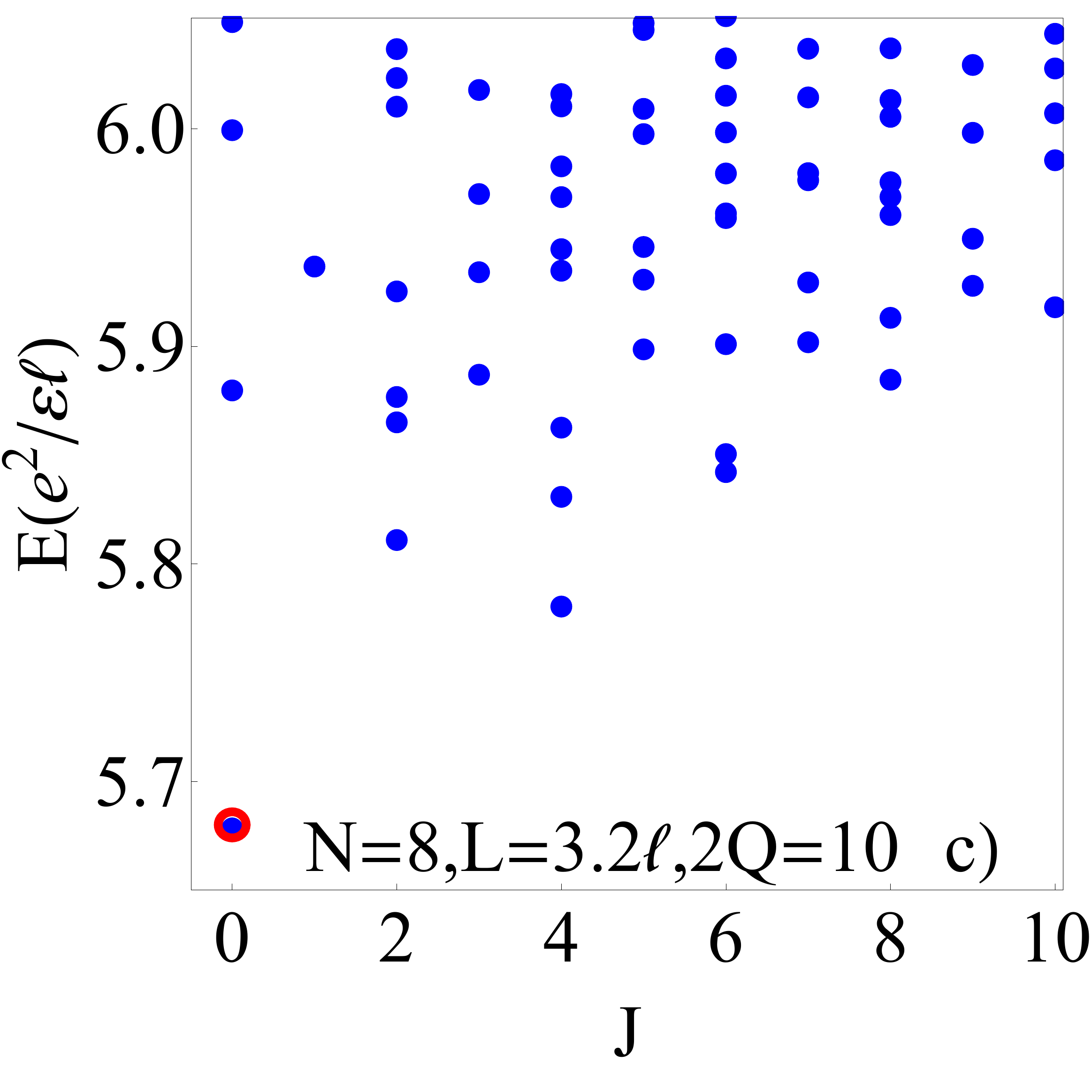},
\includegraphics[width=0.45\columnwidth]{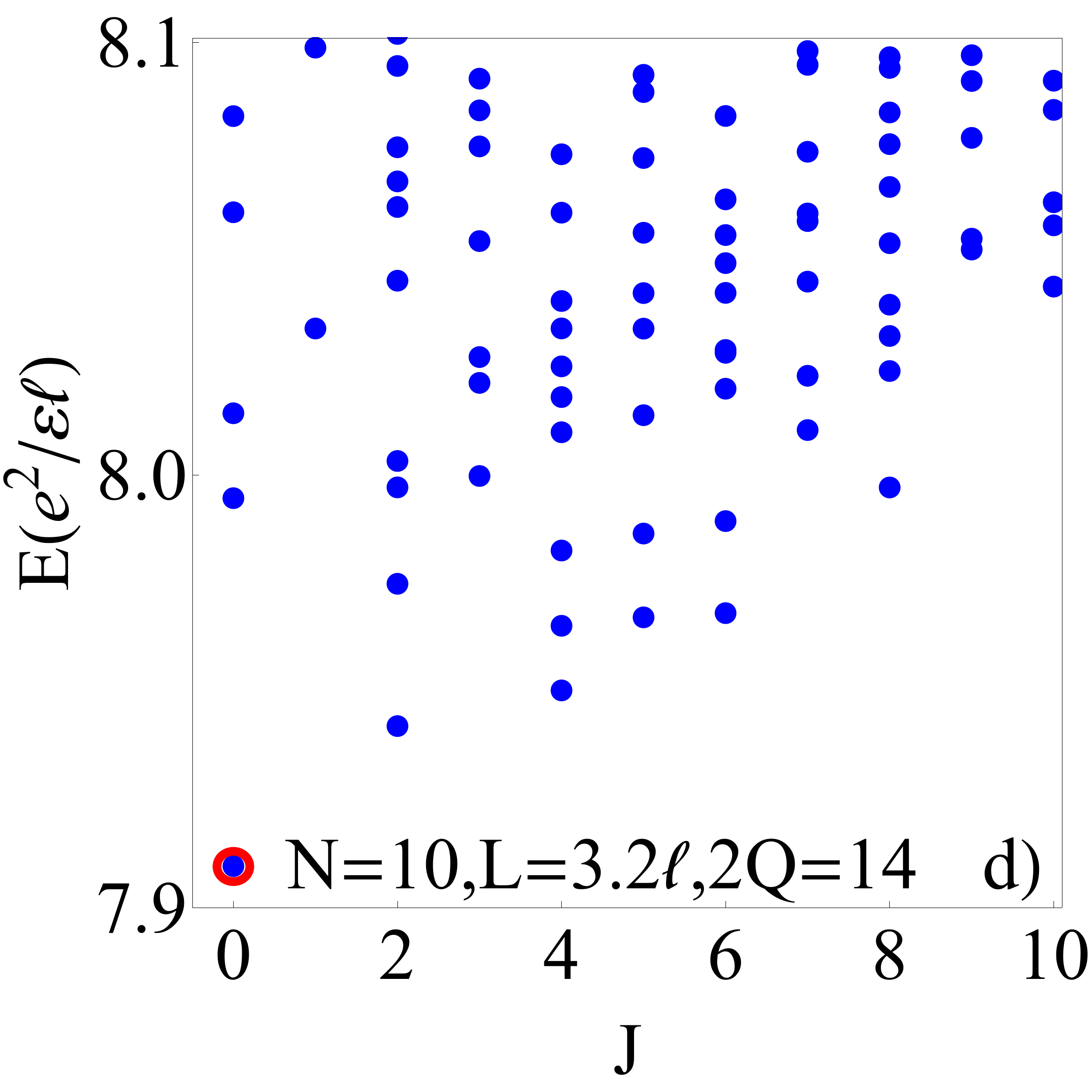}
\caption{Color online: a. LL mixing corrections to the two-hole pseudopotentials.
  Red: $\delta V({\mathcal R}=3)-\delta V({\mathcal R}=1)$; blue:
$\delta V({\mathcal R}=5)-\delta V({\mathcal R}=1)$, $w=1.6$. 
b. Three-hole irreducible pseudopotentials. 
Red: $\tilde{V}({\mathcal R_3}=5) -\tilde{V}({\mathcal R_3}=3)$;
blue:  $\tilde{V}({\mathcal R_3}=6) -\tilde{V}({\mathcal R_3}=3)$, 
$w=1.6$.  c,d. Spectra for 8 and 10 holes at $\nu=1/2$. $J=0$
ground state separated by a gap indicates an incompressible state.
} \label{fig:LLmixing}
\vspace{-0.4cm}
\end{figure}

{\it {Landau level mixing}.} The hole liquid LL mixing strength parameter $e^2/(\epsilon \ell \hbar \omega_C)$ is
very large, so we include hole virtual transitions to the other states. First, we construct a basis set with $\mathcal J$
in the two-hole state, with both holes in the same single-hole state. Holes
 undergo virtual transitions to excited levels in a
certain range of energy. A similar method was used for
electrons \cite{SimonRezayi, QuinnWootenMacek}. We diagonalize the Coulomb
interaction in this basis. The lowest energy acts
as an effective interaction. In this work, we include virtual transitions into 17 excited states that
 span the range of energy $4\hbar \omega_C$ \cite{note} due to non-regular separation between hole states.
The results are corrections $\delta
V$ to the two-hole pseudopotentials $V_0 (\mathcal {R})$. Differences between $\delta
V$ at different $\mathcal {R}$ in units of $e^4/(\ell \epsilon)^2 /(\hbar \omega_C^0)$ are shown in Fig. \ref{fig:LLmixing}a.

We next find the three-body pseudopotentials $V_{00}({\bf{\mathcal J}})$, 
${\bf{\mathcal J}}={\bf{j}}+{\bf{j}}+{\bf{j}})$ due to LL
mixing. For pseudopotentials at ${\mathcal R_3}=3j - J <9$
each value of $\mathcal{ J}$ is characterized by only one multiplet. The basis set  is made of the three-hole
states, comprised of single-hole states with energy up to $4 \hbar \omega_C$.
Using the same procedure as in the two-hole case, we find an
effective three-body pseudopotential. We then have extracted its irreducible
part $\tilde
V({\mathcal{R}}_3)$,  by subtracting the ground state energy of a three-hole system, whose interactions are given by the two-body pseudopotentials determined above.
A similar procedure was used for electrons \cite{RezayiHaldanePRB90}.
Differences between $\tilde
V$ at different $\mathcal {R}_3$ in units of $e^4/(\ell \epsilon)^2 /(\hbar \omega_C^0)$ are shown in Fig.\ref{fig:LLmixing}
b. Tables of the two- and three-body pseudopotentials in the $Q\rightarrow \infty$ limit are
given in the Supplementary Material.

{\it {$\nu=1/2$ state}.} We now consider FQHE  for $\nu=1/2$  of the ground hole state. For electrons, an incompressible  state at fillining factor $\nu$ in LL0 is obtained for a  system with $N$
particles placed on an angular momentum shell of
$2l=\nu^{-1}N+\delta$, where $\delta$ is the finite size shift given
by a topological quantum number describing the nature of
correlations \cite{Wen,MorfdAbrumenil}, and $l=Q+n$. For electrons at $\nu=1/2$,
$\delta=-3$, and this value is used to describe $\nu=5/2$ state in
the LL1 \cite{Storni}. For Luttinger holes, several Landau
indices define the spinors characterizing the two lowest states, and the results for 
electron $\delta$ \cite{Wen,MorfdAbrumenil} 
are not applicable directly. However, our simulation shows that for holes, $\delta=-3$ leads to an incompressible 
state at $n=6,8,10,12$, so that the total $j$ satisfies $2j=2N-3$ and the magnetic
monopole  is $2Q=2j-3$. The incompressible ground state persists
in the entire range $1.3<w<2.2$, that includes ground state crossings of the two
lowest $n=3$ levels shown in Fig.
\ref{fig:gen_en_spectrum}b. (We also tested that $\delta=-1$ does
not result in an incompressible states).

We first investigate whether the experimentally observed FQH state
\cite{ShayeganPRL14} is of the 331 type. The Halperin 331 state arises when there are
two species of interacting electrons, such as,
e.g., electrons in a bilayer system. The two candidates for the
degenerate species in hole FQHE are $n=3$ odd and $n=3$ even states
near and at their crossings. A translationally invariant
wavefunction of the 331 state was found in \cite{HaldaneRezayi} using
the confinement of two species of fermions to the surface of the sphere
with a monopole magnetic field in the center. Pseudopotential
describing interactions between fermions of the same species has a
repulsive character for $\mathcal R=1$ and zero for all other
$\mathcal R$. Interaction between fermions of different species is
repulsive for $\mathcal R=0$.The same construction has been
generalized for systems containing two different types of fermions,
e.g., bilayer electron liquid \cite{PetersonDasSarmaPRB}).

\begin{figure}
\centering
\includegraphics[width=0.45\columnwidth]{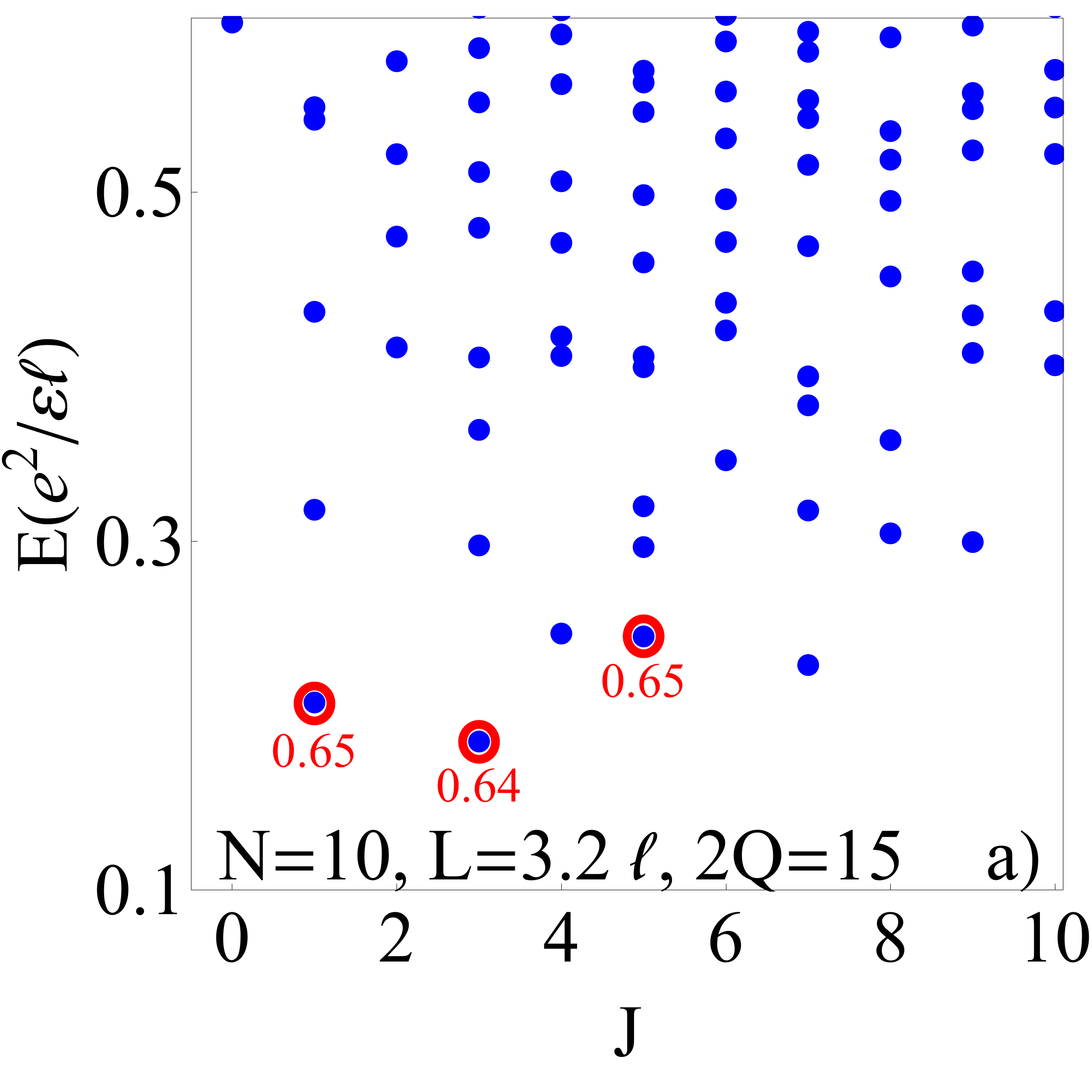},
\includegraphics[width=0.45\columnwidth]{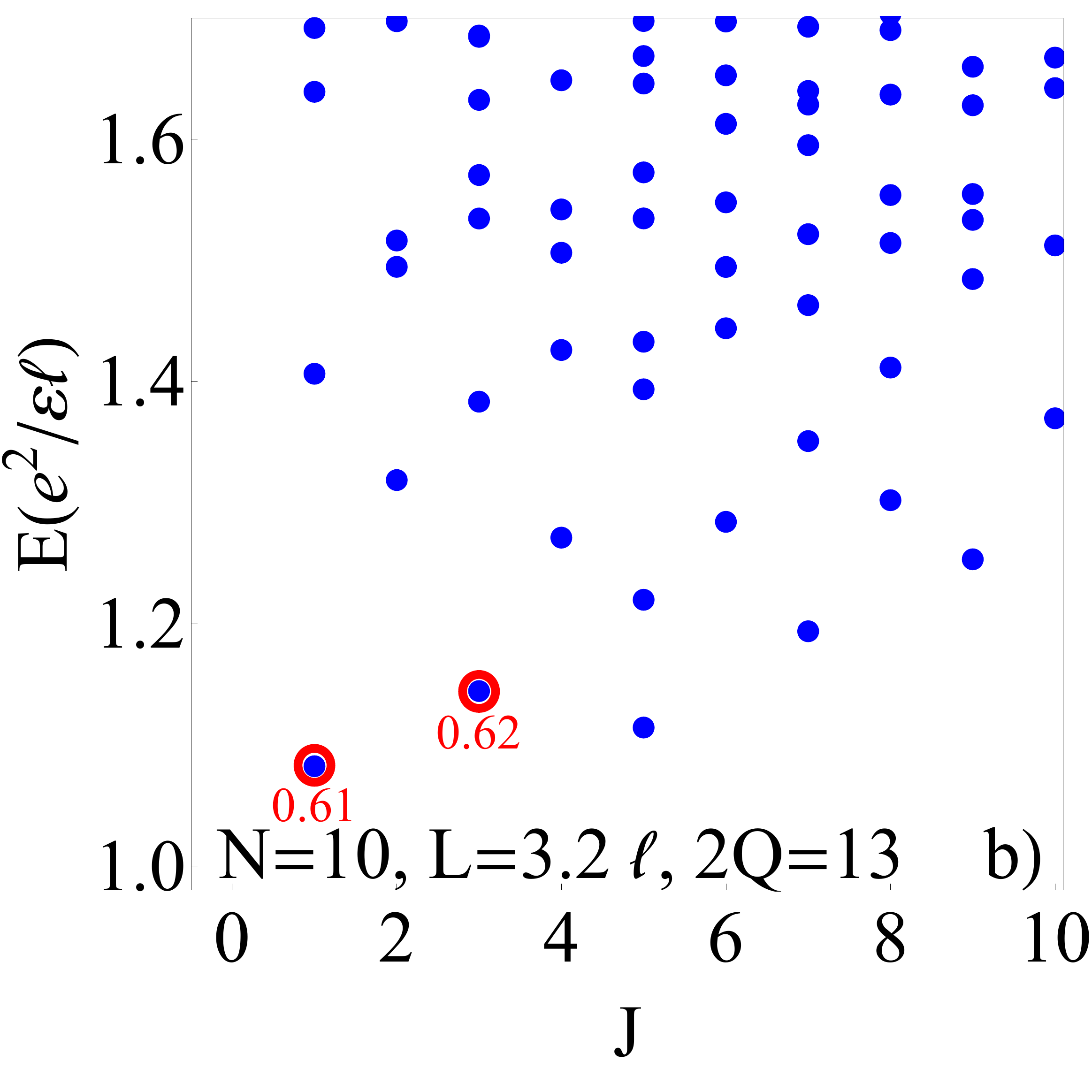}
\caption{Quasiholes (left) and quasielectron (right) pair
excitations of $\nu=1/2$  for $N=10$. Values of overlap 
between low lying excitations (red circles) and the corresponding Moore-Read excitations are
shown.} \label{fig:excitations}
\vspace{-0.4cm}
\end{figure}

Using the spherical shell configuration, we calculate the
wavefunction at $\nu=1/2$. For modeling the 331 state, the
many-hole Hilbert space must be made of the lowest states of
the double degenerate system. Its size is very large even for small
systems ($\approx 10^6$ for 10 particles). Spinor single-hole states
further complicate the simulation. Unlike the electron spin, the hole spin
is not a good quantum number. The pseudospin comprised of the
spherical shell conterparts of the planar $n=3$ odd and $n=3$ even
states is not conserved in the presence of the Coulomb interactions. Furthermore,
quantum number $J$ does not uniquely
specify a state for three holes. This makes the
simulation very challenging, and we limit it to 8 holes interacting
within the Hilbert space defined by the two crossing single-hole levels.
The exact diagonalization of the Coulomb interactions indicates that
an incompressible $J=0$ ground state is present for $w$ in the whole
range of magnetic fields that includes two crossings shown in inset of Fig. \ref{fig:gen_en_spectrum}b. \
However, the
overlap of the corresponding hole wavefunction with the 331
wavefunction \cite{HaldaneRezayi} is only $0.165-0.17$ in the whole
range of fields. It was suggested for the bilayer system \cite{Ho} that
absence of tunneling favors the 331 state. In
the present case, there is no single-particle tunnel splitting at the
crossings. However, even at crossings,
 there is significant hole-hole interactions induced
mixing of crossing levels, analog of tunneling, because of the non-conservation of the
"pseudospin" comprised of $n=3$ odd and $n=3$ even hole states.
 That precludes the possibility that the wavefunction of a many-hole system in thermodynamic limit will correspond to the Halperin 331 state.

We now consider a Moore-Read state favored by significant weight of
$u_1$  in the ground state hole spinor.
 In our case a simulation of $N=6$ and $N=12$ hole systems cannot be reliably used: besides
$\nu=1/2$, they can equally well describe filling factors
 $\nu=2/3$ and $\nu=3/5$, respectively. Systems with $N\geq 14$  holes are too large for available
computational resources, and we restrict to $N=8, 10$  holes
confined to a spherical shell. The many-body basis is built
using the hole ground state, including the LL mixing. By nature of the
spherical shell approach for holes, the effect
of a finite width of the quantum well is taken into account. The exact diagonalization
(Fig. \ref{fig:LLmixing} c-d) shows a $J=0$ ground state separated
by a gap from the continuum of states, a clear indication of an
incompressible state at $\nu=1/2$. The maximal gap occurs at $w=1.6$, very close 
to $w$  in
experiments \cite{ShayeganPRL14}. The overlap with the MR
ground state\cite{GreiterWenWilczekPRL91} at $B=10T$  is 0.8 for $N=8$ and 0.62 for
$N=10$. Excitations of FQH systems arise when flux quanta are added or subtracted.
Here adding one flux quantum in the ground state creates two quasihole excitations, and subtracting one
flux quantum gives two quasielectrons. MR quasiholes obey non-Abelian statistics \cite{MooreRead}. 
We compare excited states for $N=10$ hole system with the MR excitations, Fig.
\ref{fig:excitations}. The overlap with excitations of the MR state $\sim 0.65$,
 indicating that this FQH hole system possibly has non-Abelian statistics of excitations. 
Higher magnetic fields (at the same $w$) reduce LL mixing and enhance the MR state. At $B=16 T$, $L=200$ \AA, the overlap with MR state for $N=10$ is ~0.7.

{\it Conclusion.} We proposed the method of investigation of the  finite
size quantum Hall systems of valence band holes in a spherical shell
geometry. Our simulations show the incompressible FQH state at $\nu=1/2$  of
the ground state of holes in magnetic field.  The hole liquid at $\nu=1/2$ is not in the Halperin 331 state but
is rather described by the Moore-Read type of correlations in the many-body
ground state, with excitations having sizable overlap with
the Moore-Read Pfaffian excitations.  Experimentally, besides direct interference tests aimed at 
discovery of non-Abelian statistics \cite{NayakDasSarmaRPMQC,Review}, it is of interest to compare transport 
characteristics of  $\nu=1/2$ hole state and $\nu=5/2$ electron state in high magnetic fields.
Future work includes modeling systems
with a larger number of holes, study of FQHE at other filling factors, probing exotic states, such as the
interlayer Pfaffian \cite{Barkeshli}, and evaluation of entanglement
entropy for hole FQH systems \cite{RezayiEntang,RezayiEntangB}. This
work is supported by the U.S. Department of Energy, Office of Basic
Energy Sciences, Division of Materials Sciences and Engineering
under Award DE-SC0010544.

\vspace{-0.7cm}

\pagebreak
\widetext
\begin{center}
 \textbf{\large Supplementary Materials: \\
Non-Abelian $\nu=1/2$ quantum Hall state in $\Gamma_8$ Valence Band Hole Liquid}
\end{center}

\setcounter{equation}{0}
\setcounter{table}{0}
\setcounter{page}{1}
\makeatletter
\renewcommand{\theequation}{S\arabic{equation}}
\renewcommand{\thefigure}{S\arabic{figure}}
\renewcommand{\bibnumfmt}[1]{[S#1]}
\renewcommand{\citenumfont}[1]{S#1}
\subsection{ Differential equations for radial components
of the hole wavefunctions in a spherical shell in the presence of a magnetic monopole}

The radial envelope functions $R(r)$ are solutions of a system of coupled
differential equations
\begin{equation}
\label{eq:H_R_eq_gen}
-\gamma_1\left[\frac{1}{2}\frac{d^2}{dr^2} +
\frac{d}{dr} - \frac{l(l+1)-Q^2}{2r^2}\right]R^l_{nj}(r)
+\gamma\left[\mathcal M^2_{ll'}\frac{d^2}{dr^2}+\mathcal
M^1_{ll'}\frac{1}{r}\frac{d}{dr} +\frac{\mathcal M^0_{ll'}}{r^2}
\right]R_{nj}^{l'}(r)=E_{nj} R_{nj}^l~,
\end{equation}
with boundary conditions ($R_{nj}^l(\pm \delta R)=0$). Here matrices $ \mathcal M^i$ are given by

\begin{equation}
\label{eq:r_mat_0_matrix} \mathcal M^0=\left(
\begin{array}{cccc}
\frac{\Delta_{-\frac{3}{2}}}{2}\frac{2j-3}{2j}\left(1- \frac{3
\eta_{-\frac{1}{2}} \eta_{-\frac{3}{2}}}{2
\Delta_{-\frac{3}{2}}}\right)& \left(j-\frac{3}{2}\right)\tilde u_1&
\tilde v_1 \left(j+\frac{3}{2}\right)\left(j-\frac{1}{2}\right) &
0\\
-\left(j+\frac{1}{2}\right)\tilde u_1&
-\frac{\Delta_{-\frac{1}{2}}}{2}\frac{2j+5}{2j+2}\left(1- \frac{3
\eta_{\frac{1}{2}} \eta_{-\frac{1}{2}}}{2
\Delta_{-\frac{1}{2}}}\right)& -\tilde w \left(j-
\frac{1}{2}\right)&
\tilde v_2\left(j+\frac{1}{2}\right)\left(j+\frac{5}{2}\right)\\
\tilde v_1 \left(j-\frac{3}{2}\right)\left(j+\frac{1}{2}\right)&
\tilde w \left(j+ \frac{3}{2}\right)
&-\frac{\Delta_{\frac{1}{2}}}{2}\frac{2j-3}{2j}\left(1-
\frac{3 \eta_{\frac{3}{2}} \eta_{\frac{1}{2}}}{2 \Delta_{\frac{1}{2}}}\right) &-\tilde u_2 \left(j+ \frac{1}{2}\right) \\
0&\tilde v_2 \left(j-\frac{1}{2}\right)\left(j+\frac{3}{2}\right)
&\tilde u_2 \left(j+ \frac{5}{2}\right)
&\frac{\Delta_{\frac{3}{2}}}{2}\frac{2j+5}{2j+2}\left(1- \frac{3
\eta_{\frac{5}{2}} \eta_{\frac{3}{2}}}{2
\Delta_{\frac{3}{2}}}\right)
\end{array}\right),
\end{equation}

\begin{equation}
\label{eq:r_mat_1_matrix} \mathcal M^1=\left(
\begin{array}{cccc}
-\frac{2j-3}{2j} \left(1-3\eta_{-\frac{3}{2}}
\eta_{-\frac{1}{2}}\right)& \left(2j+3\right)\tilde u_1& 2 \tilde
v_1 \left(j+1\right) &
0\\
-\left(2j-5\right)\tilde u_1& \frac{2j+5}{2j+2}
\left(1-3\eta_{-\frac{1}{2}} \eta_{\frac{1}{2}}\right)& -\tilde w
\left(2 j+5\right)&
2\tilde v_2\left(j+2\right)\\
-2 \tilde v_1 \left(j-1\right)& \tilde w \left(2 j-3\right) &
\frac{2j-3}{2j} \left(1-3\eta_{\frac{3}{2}}
\eta_{\frac{1}{2}}\right)&
-\tilde u_2 \left(2 j+7\right) \\
0&-2\tilde v_2 j &\tilde u_2 \left(2j-1\right) & -\frac{2j+5}{2j+2}
\left(1-3\eta_{\frac{3}{2}} \eta_{\frac{5}{2}}\right)
\end{array}\right),
\end{equation}

\begin{equation}
\label{eq:r_mat_2_matrix} \mathcal M^2=\left(
\begin{array}{cccc}
-\frac{2j-3}{4j} \left(1-3\eta_{-\frac{3}{2}}
\eta_{-\frac{1}{2}}\right)& 2 \tilde u_1& \tilde v_1 &
0\\
2\tilde u_1& \frac{2j+5}{4j+4} \left(1-3\eta_{-\frac{1}{2}}
\eta_{\frac{1}{2}}\right)& -2\tilde w &
\tilde v_2\\
\tilde v_1& -2\tilde w  & \frac{2j-3}{4j}
\left(1-3\eta_{\frac{3}{2}} \eta_{\frac{1}{2}}\right)&
-2\tilde u_2  \\
0&\tilde v_2  &-2 \tilde u_2& -\frac{2j+5}{4j+4}
\left(1-3\eta_{\frac{3}{2}} \eta_{\frac{5}{2}}\right)
\end{array}\right),
\end{equation}
where $\eta_k=s/(j+k)$, $\Delta_k=(j+k)(j+k+1)-Q^2$ and
\begin{eqnarray}
\tilde u_1&=&\eta_{\frac{1}{2}}\sqrt{\frac{3 (1+j)}{4
j}}\sqrt{1-\eta_{-\frac{1}{2}}^2}~,\\
\tilde u_2&=&\eta_{\frac{1}{2}}\sqrt{\frac{3 j}{4
(j+1)}}\sqrt{1-\eta_{\frac{3}{2}}^2}~,\\
\tilde v_1&=&\frac{\sqrt{3}}{2 j}\sqrt{\left(j-\frac{1}{2}\right)
\left(j+\frac{3}{2}\right)}
\sqrt{\left(1-\eta_{-\frac{1}{2}}^2\right)\left(1-\eta_{\frac{1}{2}}^2\right)}~,\\
\tilde v_2&=&\frac{\sqrt{3}}{2(1+j)}\sqrt{\left(j-\frac{1}{2}\right)
\left(j+\frac{3}{2}\right)}
\sqrt{\left(1-\eta_{\frac{1}{2}}^2\right)\left(1-\eta_{\frac{3}{2}}^2\right)}~,\\
\tilde w&=& \frac{3}{2}
\sqrt{\frac{1}{j(j+1)}}\sqrt{\eta_{\frac{3}{2}} \eta_{-\frac{1}{2}}
\left(1-\eta_{\frac{1}{2}}^2\right)}
\end{eqnarray}

Taking the limit of $Q \rightarrow \infty$ , and keeping $R\sqrt{Q}=w$, $j=Q-3/2+n$,
after some algebraic transformations, we re-write
Eq. (\ref{eq:H_R_eq_gen}) in the following form:
\begin{equation}
\hat M \left(
\begin{array}{c}
rR_1(r)\\
i r R_2(r)\\
-r R_3(r) \\
-i r R_4(r)
\end{array}
\right)=E \left(
\begin{array}{c}
rR_1(r)\\
i r R_2(r)\\
-r R_3(r) \\
-i r R_4(r)
\end{array}
\right)~,
\end{equation}
with the matrix operator

\begin{equation}
\hat M=\left(
\begin{array}{cccc}
-\frac{1}{2m_h}\frac{\partial^2}{\partial r^2}+\gamma_+(n-1)-
\frac{3\gamma}{2}
&-i \gamma \sqrt{6(n-2)} \frac{\partial}{\partial r}  & -\gamma\sqrt{3(n-1)(n-2)} &0\\
-i \gamma \sqrt{6(n-2)} \frac{\partial}{\partial r}
&-\frac{1}{2m_l}\frac{\partial^2}{\partial r^2}+\gamma_-
n+ \frac{3\gamma}{2} &0 & -\gamma \sqrt{3 n (n-1)}\\
-\gamma \sqrt{3 (n-2) (n-1)} &0 &
-\frac{1}{2m_l}\frac{\partial^2}{\partial r^2}+\gamma_-
(n+1)-\frac{3 \gamma}{2} & -i\gamma
\sqrt{6n}\frac{\partial}{ \partial r}\\
0&-\gamma \sqrt{3n(n-1)}&-i\gamma\sqrt{6n}\frac{\partial}{\partial
r}&-\frac{1}{2m_h}\frac{\partial^2}{\partial
r^2}+\gamma_+(n+2)-\frac{3\gamma}{2}
\end{array}
\right)~,
\end{equation}
where $\gamma_{\pm}=\gamma_1\pm\gamma$,
$m_h=(\gamma_1-2\gamma)^{-1}$ and $m_l=(\gamma_1+2\gamma)^{-1}$.
This is exactly the equation for the envelope $z$-functions in a planar
geometry. Thus, we have shown that in the limit of very large radius and
monopole limit ($R, Q \rightarrow \infty$ with $R/\sqrt{Q}= w$) the wavefunctions in a spherical shell
converge to the planar limit.

\subsection{Zeemann term in a spherical shell geometry}

We include the pure Zeeman term ${\cal{H}_Z}=\kappa \mathbf{s}\cdot
\mathbf{H}/|H|$ of the Luttinger Hamiltonian that is due to direct
coupling of the spin 3/2 of holes  with a magnetic field, where energy
is in the units of the free electron cyclotron energy and $\kappa$ is
the Luttinger constant. For a spherical shell, a magnetic
field is not a constant in the radial direction. However, we evaluate Zeeman
term approximating its value by taking the magnetic field value on the sphere of radius $R$,
half way between the inner and outer spherical boundaries of the shell. This choice recovers the planar Zeeman term  in the large $Q$ limit.
The Zeemann term is diagonal in the angular momentum
representation and its eigenvalues correspond to spin projections
$\pm 1/2$ , $\pm 3/2$ in the limit of large $Q$. The angular part of
the wavefunction is
\begin{eqnarray} \label{eq:ang_part}
&\psi_{Qjm}(\theta, \phi)=\left(
\begin{array}{c}
\left<j,m\right|\left.l,m-\frac{3}{2};\frac{3}{2},+\frac{3}{2}\right> Y_{Q,l,m-\frac{3}{2}}(\theta, \phi)\\
\left<j,m\right|\left.l,m-\frac{1}{2};\frac{3}{2},+\frac{1}{2}\right> Y_{Q,l,m-\frac{1}{2}}(\theta, \phi)\\
\left<j,m\right|\left.l,m+\frac{1}{2};\frac{3}{2},-\frac{1}{2}\right> Y_{Q,l,m+\frac{1}{2}}(\theta, \phi)\\
\left<j,m\right|\left.l,m+\frac{3}{2};\frac{3}{2},-\frac{3}{2}\right>
Y_{Q,l,m+\frac{3}{2}}(\theta, \phi)
\end{array}
\right)~,
\end{eqnarray}
where $\left<j,m_j\right| \left.l,m-{l};\frac{3}{2},m_s\right>$ are
the Clebsch-Gordan coefficients of $\mathcal
{\bf{j}}={\bf{l}}+{\bf{s}}$. In the presence of a radial magnetic field,  ${\cal{H}_Z}=\kappa s_r$, where $s_r$ is the radial component of $3/2$ spin matrix, and the non-zero matrix elements of the Zeemann interaction are
\begin{eqnarray}
<\psi_{Q,l+3/2,m}|\kappa s_r| \psi_{Q,l+3/2,m}>&=&3\kappa(j-3/2)/2Q,\nonumber\\
<\psi_{Q,l+1/2,m}|\kappa s_r| \psi_{Q,l+1/2,m}>&=&\kappa(j-7/2)/2Q,\nonumber\\
<\psi_{Q,l-1/2,m}|\kappa s_r| \psi_{Q,l-1/2,m}>&=&-\kappa(j+9/2)/2Q,\nonumber\\
<\psi_{Q,l-3/2,m}|\kappa s_r|
\psi_{Q,l-3/2,m}>&=&-3\kappa(j+5/2)/2Q.
\end{eqnarray}

\subsection{Matrix elements of hole-hole Coulomb interactions}

We derive the matrix elements of the hole-hole
interactions by  using the angular momentum addition and the
 relation \cite{jacksonbook}:
\begin{equation}
\frac{1}{|{\bf{r_1}}-{\bf{r_2}}|}=
4\pi\sum_{k=0}^{\infty}\sum_{\mu=-k}^k\frac{1}{2k+1}\frac{r_{<}^k}{r_{>}^{k+1}}
Y_{k\mu}^*(\theta',\phi')Y_{k\mu}(\theta,\phi)~,
\end{equation}
where ${\bf{r}}_{2}$, $r_{</>}$ being the smaller (or bigger) of
$r_1$ and $r_2$, and $Y_{k\mu}(\theta,\phi)$ is a spherical function, which is a monopole harmonics function
$Y_{Q=0,k,\mu}(\theta,\phi)$ . A straightforward
calculation by using the integral of the three monopole harmonic functions (Eq.
(1) of [\onlinecite{MonopoleHarmonics2}])   leads to:

\begin{eqnarray}
\label{eq:mat_el_final} \braket{Q; \alpha_1',j_1,'m_1-\eta;
\alpha_2',j_2',m_2+\eta\left|\frac{e^2}{\epsilon|{\bf{r_1}}-{\bf{r_2}}|}\right|Q;
\alpha_1,j_1,m_1;
\alpha_2,j_2,m_2}=\frac{e^2}{\epsilon}\sum_{\mu_1',\mu_2'}
\sum_{\mu_1,\mu_2}\sum_{\sigma_1,\sigma_2}\sum_{k=0}^{\infty}
(-1)^{\eta}\times~~~~~\nonumber\\
\braket{  j_1'+\mu_1,m_1-\sigma_1-\eta;\frac{3}{2}, \sigma_1 |
j_1',m_1-\eta } \braket{ j_2'+\mu_2,m_2+\eta-\sigma_2;\frac{3}{2},
\sigma_2 | j_2',m_2+\eta
}\braket{  j_1+\mu_1,m_1-\sigma_1;\frac{3}{2}, \sigma_1 | j_1,m_1 
}\nonumber\\
\braket{ j_2+\mu_2,m_2-\sigma_2;\frac{3}{2}, \sigma_2 | j_2',m_2 }
\braket{k,0;j_1+\mu_1,Q|j_1'+\mu_1',Q} 
\braket{j_1'+\mu_1',\sigma_1+\eta-m_1;k,-\eta|j_1+\mu_1;-m_1+\sigma_1}~~~~\nonumber\\
\braket{k,0;j_2+\mu_2,Q|j_2'+\mu_2',Q}\nonumber\braket{j_2'+\mu_2',\sigma_2+\eta-m_2;k,-\eta|j_2+\mu_2;-m_2+\sigma_2}
\int_{\sqrt{Q}-w}^{\sqrt{Q}+w} \int_{\sqrt{Q}-w}^{\sqrt{Q}+w}  dr_1
dr_2 r_1^2  r_2^2 \frac{r_<^k}{r_>^{k+1}}\nonumber\\
R_{j_1'}^{j_1'+\mu_1'}(r_1) R_{j_1}^{j_1+\mu_1}(r_1)
R_{j_2'}^{j_2'+\mu_2'}(r_2) R_{j_2}^{j_2+\mu_2}(r_2).
~~~~~~~~~~~~~~~~~~~~~~~~~~~~~~~~~~~~~~~~~~~~~~~~~~~~~~~~~~~~~~~~~~~~~~~~~~~~~~~~~~~~~~~
\end{eqnarray}
Evaluating this equation, we use explicitly known expressions for the Clebsch-Gordan coefficients,
and numerically calculate the integrals by applying Simpson rule for evaluating the numerical quadrature involving
radial wavefunctions.

\subsection{Two- and three-body pseudopotentials of the hole-hole
interaction}

\begin{table}[h]
\caption{Corrections due to LL mixing to the ground state two-body
pseudopotentials in units of $e^4/\varepsilon^2 \ell^2 \hbar
\omega_C^{0}$} 
\vspace{0.6cm}
\centering
\begin{tabular}{c | c | c | c | c |c |c |c}
$L/2\ell$ & 1.4 & 1.5 & 1.6 & 1.7 & 1.8 & 1.9 & 2.0\\
\hline &&&&&&&\\
$\delta V_5-\delta V_3$ & 0.0248 & 0.0211 & 0.0187 &0.0168 & 0.0153 & 0.0140 & 0.0129 \\
\hline &&&&&&&\\
$\delta V_7-\delta V_3$& 0.0314 &0.0266 & 0.0235 & 0.0212 & 0.0192 &0.0176& 0.0161\\
\end{tabular}
\end{table}

\begin{table}[h]
\caption{The ground state irreducible three-body pseudopotentials in
units of $e^4/\varepsilon^2 \ell^2 \hbar \omega_C^{0}$} 
\vspace{0.6cm}
\centering
\begin{tabular}{c | c | c | c | c |c |c |c}
$L/2\ell$ & 1.4 & 1.5 & 1.6 & 1.7 & 1.8 & 1.9 & 2.0\\
\hline &&&&&&&\\ $\tilde V_5-\tilde V_3$ & 0.0015 & 0.0108 & 0.0150
& 0.0174 &
0.0188 & 0.0196 & 0.0202\\  \hline &&&&&&&\\
$\tilde V_6-\tilde V_3$ & -0.0357& -0.0240 & -0.0177 & -0.0136 & -0.0107 & -0.0086 & -0.0070\\
\end{tabular}
\end{table}


\end{document}